\documentclass[conference]{IEEEtran}
\usepackage{array}
\usepackage{booktabs} \usepackage{amssymb,latexsym,amsmath}     \usepackage{booktabs}
\usepackage{multirow}
\usepackage{booktabs, multicol, multirow}
\usepackage{graphicx}
\usepackage{epstopdf}
\usepackage{tikz}
\usepackage{url}
\usepackage{cite}
\usepackage{subfig}
\usepackage{flushend}
\usepackage{color, colortbl}
\definecolor{LRed}{rgb}{1,.8,.8}
\definecolor{LGreen}{rgb}{0.8,1,0.8}
\definecolor{HRed}{rgb}{1,.2,.2}
\definecolor{Yellow}{cmyk}{0,0,0.5,0}
\usepackage{array}
 \usepackage{booktabs} \usepackage{bigstrut}
\makeatletter
\newlength\mylena
\newlength\mylenb
\newcommand\mystrut[1][2]{    \setlength\mylena{#1\ht\@arstrutbox}    \setlength\mylenb{#1\dp\@arstrutbox}    \rule[\mylenb]{0pt}{\mylena}}
\makeatother
\usepackage{colortbl}
\usepackage{hhline}

\usepackage{bigstrut}
\newlength{\Oldarrayrulewidth}

\usepackage{tabu}

\def \sys {\textit{MonoSense}}
\IEEEoverridecommandlockouts

\begin{document}

\title{\Large The Diversity and Scale Matter:\\Ubiquitous Transportation Mode Detection using Single Cell Tower Information}

\author{
\IEEEauthorblockN{Ali Mohamed AbdelAziz}
\IEEEauthorblockA{Wireless Research Center\\Egypt-Japan Univ. of Science and Tech. (E-JUST)\\
Alexandria, Egypt\\
Email: ali.abdelgalil@ejust.edu.eg}
\and
\IEEEauthorblockN{Moustafa Youssef}
\IEEEauthorblockA{Wireless Research Center\\Egypt-Japan Univ. of Science and Tech. (E-JUST)\\
Alexandria, Egypt\\
Email: moustafa.youssef@ejust.edu.eg}
\thanks{Moustafa Youssef is currently on sabbatical from Alexandria University, Egypt.}
}

\maketitle

\begin{abstract}
Detecting the transportation mode of a user is important for a wide range of applications. While a number of recent systems addressed the transportation mode detection problem using the ubiquitous mobile phones, these studies either leverage GPS,  the inertial sensors, and/or multiple cell towers information. However, these different phone sensors have high energy consumption, limited to a small subset of phones (e.g. high-end phones or phones that support neighbouring cell tower information), cannot work in certain areas (e.g. inside tunnels for GPS), and/or work only from the user side.

In this paper, we present a transportation mode detection system, \sys{}, that leverages the phone serving cell information \textbf{only}. The basic idea is that the phone speed can be correlated with features extracted from both the serving cell tower ID and the received signal strength from it. To achieve high detection accuracy with this limited information, \sys{} leverages diversity along multiple axes to extract novel features. Specifically, \sys{} extracts features from both the time and frequency domain information available from the serving cell tower over different sliding widow sizes. More importantly, we show also that both the logarithmic and linear RSS scales can provide different information about the movement of a phone, further enriching the feature space and leading to higher accuracy.

Evaluation of \sys{} using $135$ hours of cellular traces covering 485 km and collected by four users using different Android phones shows that it can achieve an average precision and recall of 89.26\% and 89.84\% respectively in differentiating between the stationary, walking, and driving modes using only the serving cell tower information, highlighting \sys{} ability to enable a wide set of intelligent transportation applications.

\end{abstract}
\IEEEpeerreviewmaketitle

\section{Introduction}
\graphicspath{ {./} }

The inference  of the human transportation mode (e.g. walking, driving, etc) is important for a wide range of applications such as human behavior monitoring \cite{consolvo2008activity}, road traffic estimation \cite{snapnet}, evaluating transportation related measures and policies \cite{mohan2008nericell,aly_map14,sheikh2014demonstrating}, among others. During the last decade, there has been rapid growth in the sensing capabilities of commodity phones combined with their ease of programming and large market penetration rate. Therefore, a number of unobtrusive systems that leverage the phone different sensors have been proposed  \cite{stenneth11,widhalm2012transport,lu2010jigsaw,bolbol2012inferring,miluzzo2008sensing,Wang2010,yang2009toward}. Specifically, GPS-based systems \cite{stenneth11,widhalm2012transport,lu2010jigsaw,bolbol2012inferring} that depend on the GPS location or its alternatives \cite{ibrahim2011hidden,cellsense,ibrahim2010cellsense,ibrahim2013enabling,aly2013dejavu,GAC}, inertial-sensors (mainly the accelerometer) \cite{miluzzo2008sensing,Wang2010,yang2009toward}, or a combination of both can provide different accuracy for transportation mode detection with different garrulities. However, these systems suffer from high energy consumption, do not work everywhere (e.g. GPS does not work in tunnels and urban canyons), work only on smart phones, work in special phone positions (e.g. outside pocket with clear line of sight to the satellites or in hand) and do not work from the cellular provider side.

To address these issues, a number of systems have been recently introduced that leverage cellular network information only (\emph{multiple} cell towers IDs and received signal strength (RSS) from them), e.g. \cite{anderson2006practical,Sohn06,reddy10,Kassem:VTC12,al2012rf}, to detect whether a user is still, walking, or in a motorized transport. Nevertheless, previous cellular data based systems require the information from the serving cell tower as well as \textbf{\emph{all neighboring cell towers information}}. Given the fact that the majority of Android phones, which account for more than 80\% of the smart phone market, \textbf{\emph{only provide the serving cell tower information }}\cite{webstatandroid2}, this sparks the  need for new methods that can detect the mode of transportation accurately, with only a single cell tower information, from both the phone or cellular provider side, and work with any phone (including low end phones).

In this paper, we propose \sys{}, a system that leverages the ID and RSS information from the serving cell tower \textbf{\emph{only}} from any commodity cellular phone to differentiate between three human modes of transportation: Stationary, walking, and driving. The basic idea is that different modes of transportation can be mapped to different speeds. These speeds in turn, can be correlated with features extracted from the serving cell tower ID and RSS. To address the confusion among different transportation modes with the limited available information, \sys{} draws on two main concepts: (a) a novel interesting observation about the input RSS and (b) features diversity. For the former, we show that both the linear and logarithmic scales of the RSS can provide different information about the movement pattern, leading to more accurate differentiating between the different transportation modes. For the latter, the diversity in the RSS spaces is combined with the diversity of features extracted from both the time and frequency domains as well as diversity of the window sizes the features are extracted from, expanding the space of available features and providing a better possibility for removing the ambiguity between classes.

We evaluate \sys{} using real-world data collected by four persons using different Android phones  with different cellular  operators over a period of eight months covering $135$ hours of cellular traces. Our results show that \sys{} can achieve an average precision and recall of 89.26\% and 89.84\% respectively in differentiating between the stationary, walking, and driving modes using only the serving cell tower information

In summary, our contribution in this paper is three-fold:
\begin{itemize}
  \item We provide the architecture and details of the \sys{} system that can provide accurate, ubiquitous, and energy-efficient transportation mode detection using the information from the \textbf{\emph{serving cell tower only}}.
  \item We show that the RSS linear scale can provide independent information about the RSS in addition to the RSS logarithmic scale, which is the \textbf{\emph{only scale}} used in traditional cellular-based transportation mode detection systems. This extends the features pool and helps in handling the limited information available for \sys{}.
  \item We implement and evaluate \sys{} in typical environments with different phones, operators, and users.
\end{itemize}

 The remainder of this paper is organized as follows.  Section~\ref{sec:system} presents the \sys{} system architecture as well as details of the different system components.
 We evaluate the system in Section~\ref{sec:evaluation} and conclude in Section~\ref{Sec:conclusion}.

    	                                                                \section{The \sys{} System}\label{sec:system}
In this section, we start by providing an overview of \sys{} architecture and the principal of operation followed by the details of the different system components, mainly: preprocessing, features extraction, differences between the RSS logarithmic and linear scales, and the classifier used. We end the section by a discussion of different aspects of \sys{}.

\subsection{Overview}
\begin{figure}[!t]
	\centering
		\includegraphics[width=1.0\linewidth]{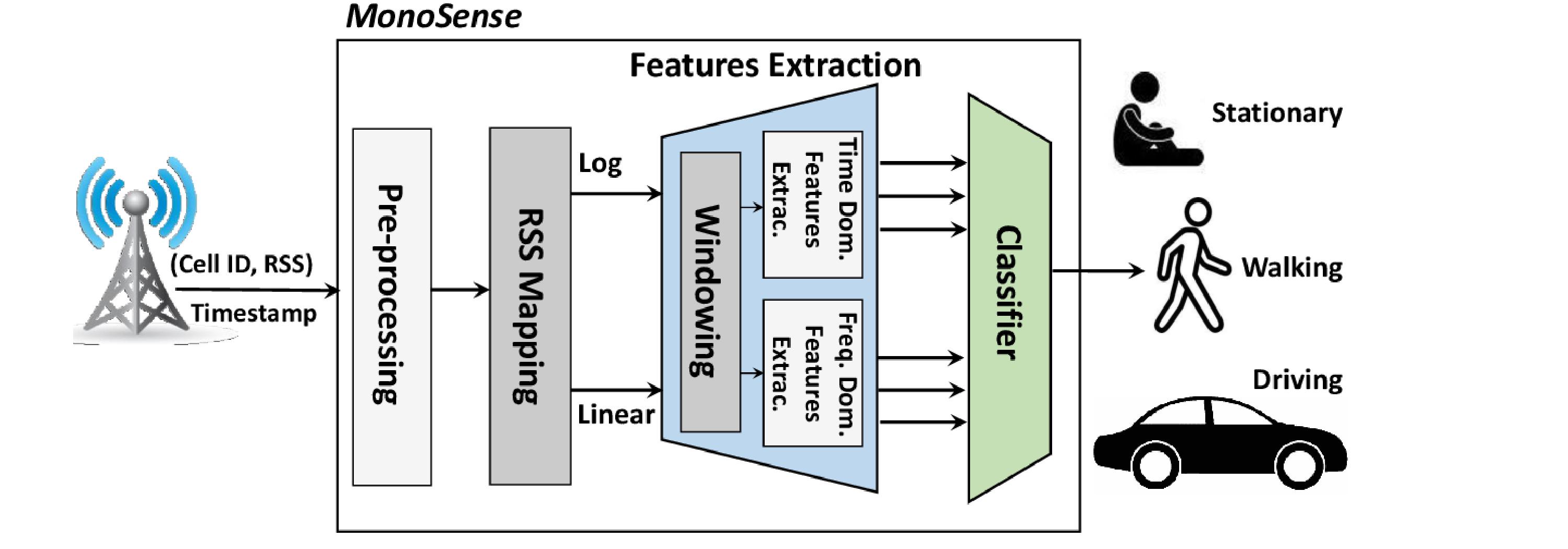}
	\caption{\sys{} system architecture.}
	\label{fig:Arc}
\end{figure}

\begin{figure*}[!t]
  \centering
  \subfloat[Number of unique cell IDs]{\label{fig:handoff}\includegraphics[width=50mm]{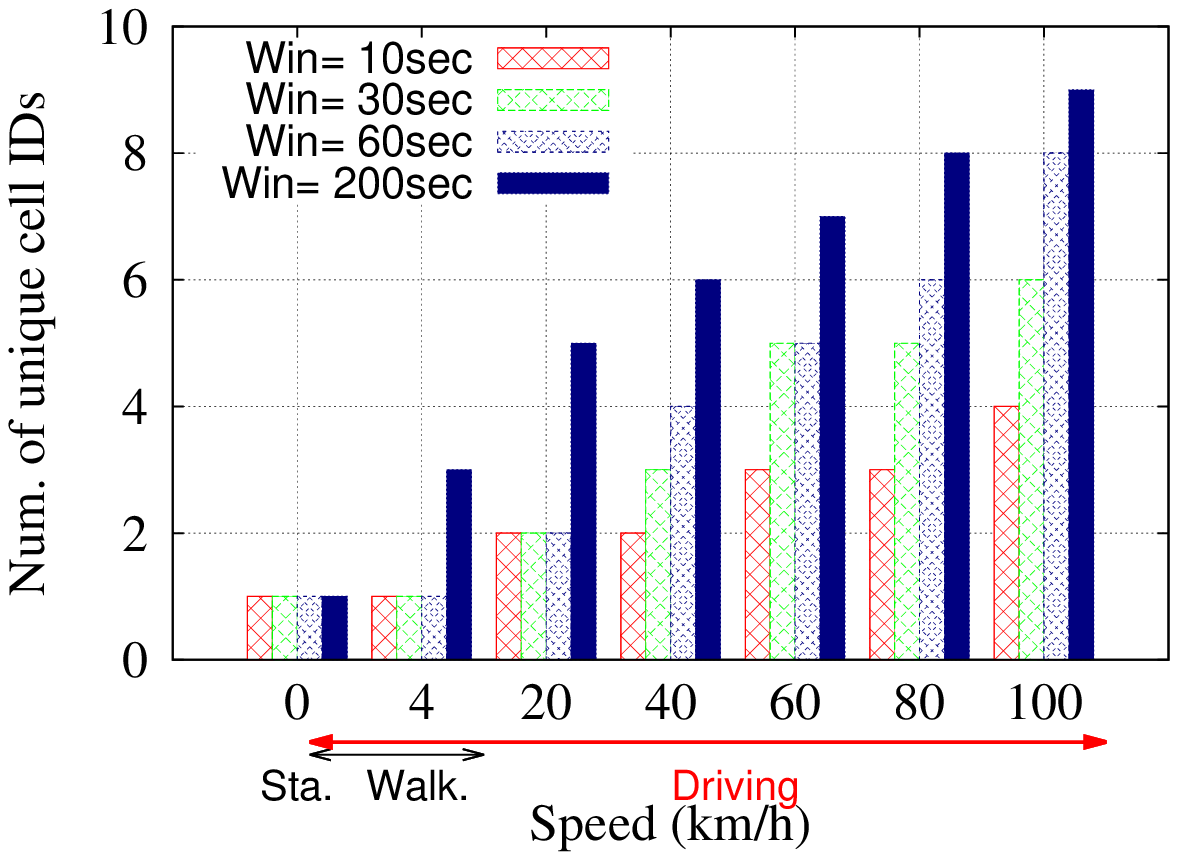}}
  \subfloat[Cell residence time]{\label{fig:ResisdenceTime}\includegraphics[width=50mm]{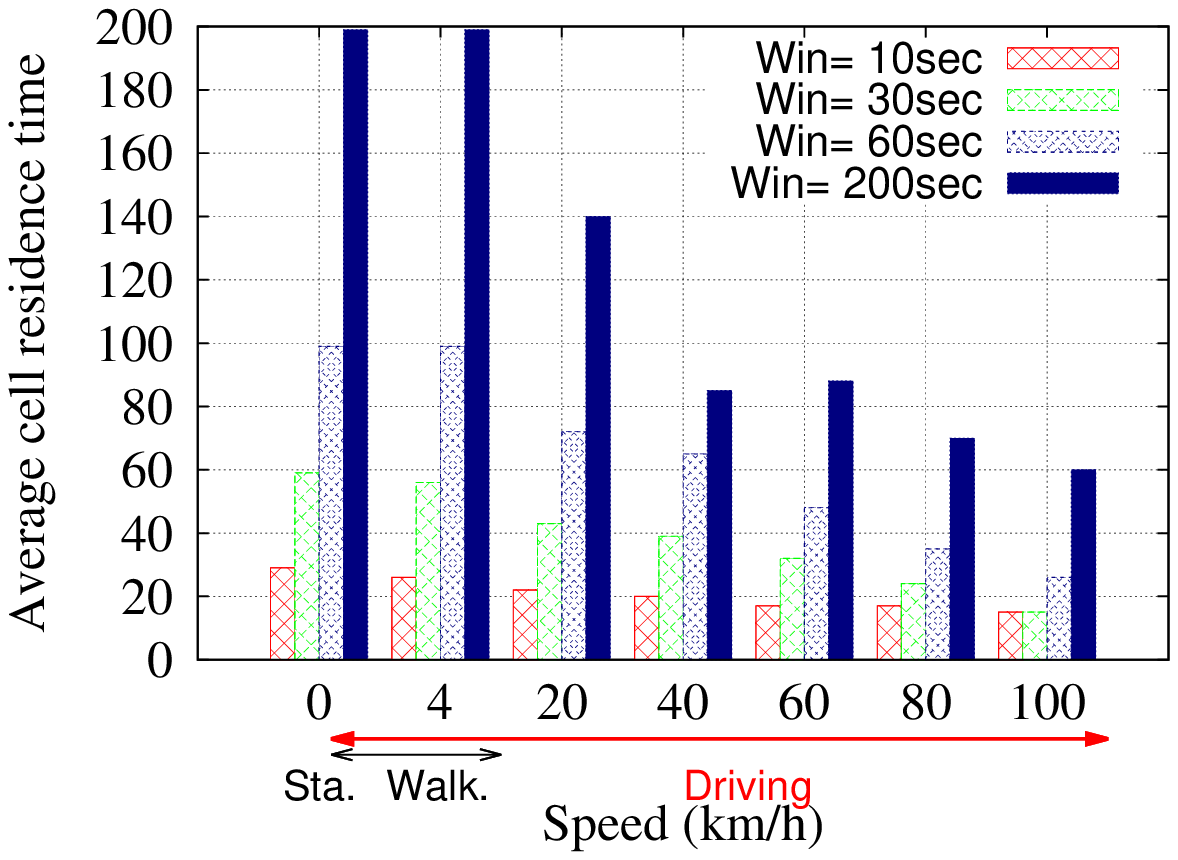}}
	  \subfloat[Average RSS difference between consecutive measurements]{\label{fig:AvgEUC}\includegraphics[width=50mm]{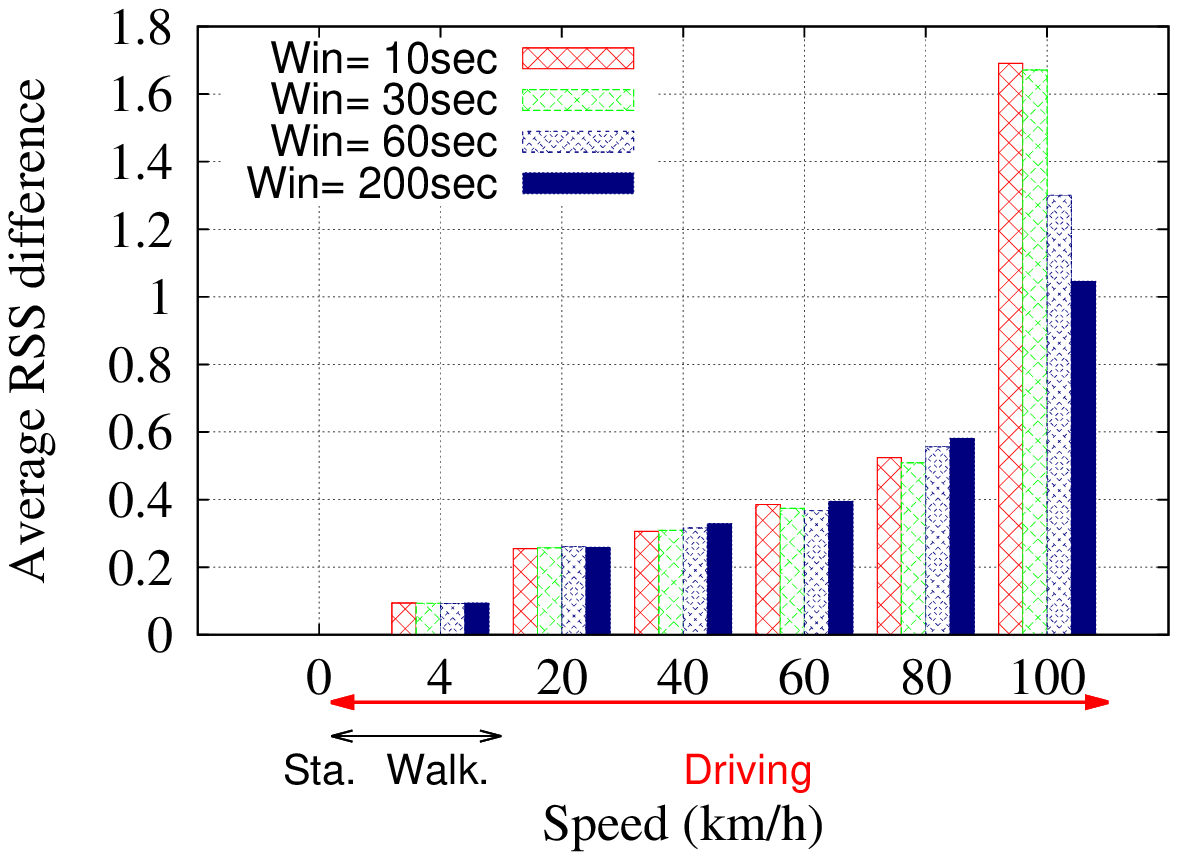}}
  \subfloat[Variance of RSS]{\label{fig:rss_var}\includegraphics[width=50mm]{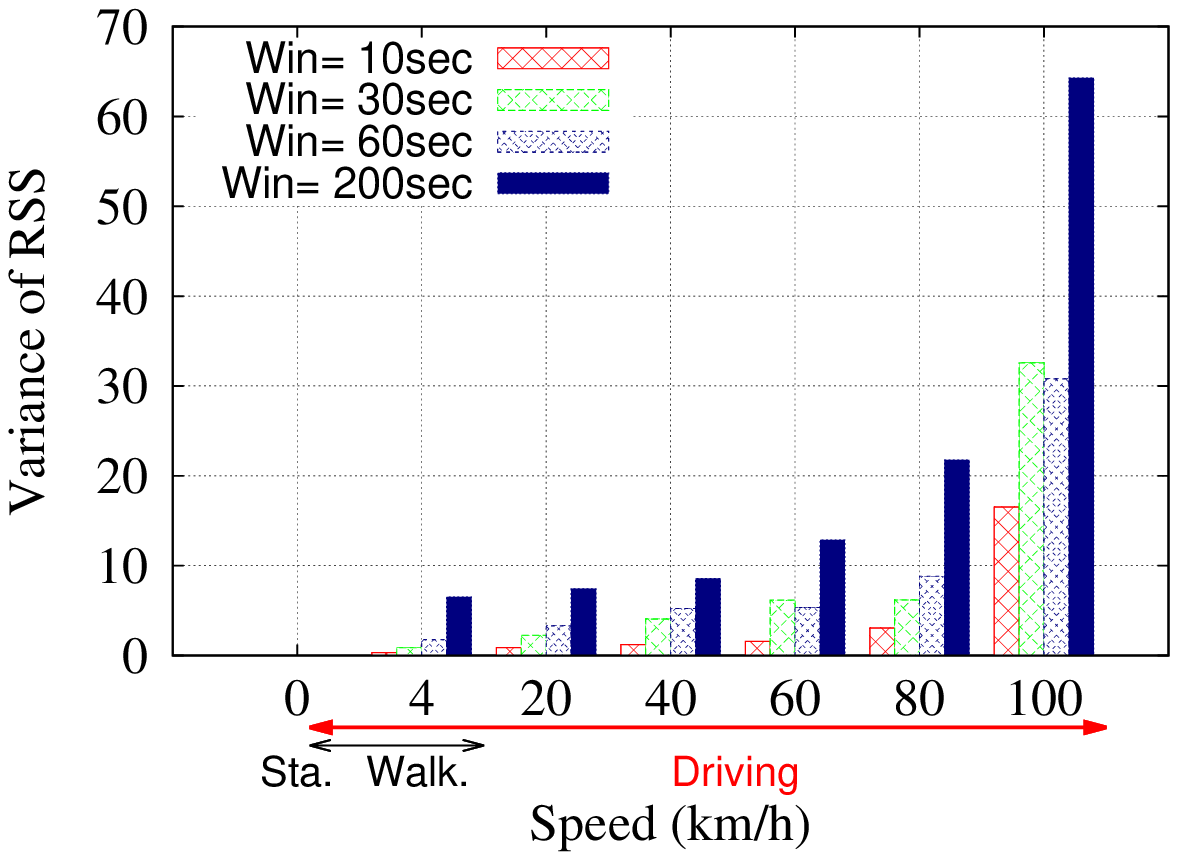}}

		\caption{Effect of the speed on the different time domain features. All features extracted within a certain window size.}
\label{fig:basic_oper}
\end{figure*}

Figure~\ref{fig:Arc} shows the system architecture. The only information available for \sys{} to detect the mode of transportation is the serving cell ID and associated RSS. The basic idea that \sys{} builds on is that some features extracted from these two sources of data can be correlated with the user speed. For example, a phone in a fast moving car would encounter faster changes in the RSS and see more changes in the associated cell towers compared to a stationary phone (within the same time duration). This maps to a higher difference between adjacent RSS readings, higher variance in RSS, and higher handoff (changes in the serving cell tower ID) frequency as shown in Figure~\ref{fig:basic_oper}. However, due to the limited information available from the serving cell tower only, \sys{} diversifies its features pool to increase the classification accuracy including using features from the linear and logarithmic RSS space, in time and frequency domains, and over different RSS window sizes.

The system starts by collecting a stream of serving cell tower information ($d=d_0, d_1, ...)$, where each $d_i$ is an ordered pair $(\textrm{ID}_i, \textrm{RSS}_i)$ representing the ID of the serving cell tower and the associated received signal strength (RSS) in logarithmic scale (this is the default scale used by the Android API) at sample $i$. These samples can either be collected from the cell phone or the cellular provider side. For the cell phone case, since this information is available from almost all phones and during the normal phone operation, \sys{} consumes \textbf{\emph{zero extra energy}}, making it a \emph{ubiquitous and energy-efficient solution}.

The input data stream is then pre-processed to filter noisy data and then the RSS is mapped from the logarithmic to the linear scale. It then passed to a feature extractor to extract different features in the time and frequency domain over different sliding window sizes to enrich the feature space.

A decision tree classifier is then applied to differentiate between the three modes of transportation, corresponding to three different ranges of speed: stationary (speed almost zero km/h), walking (maximum speed about 3-6 km/h, and in vehicle (free running speeds from 40-100 km/h). Note that the different modes may \textbf{\emph{overlap}} in some cases. For example, in traffic congestion, the car speed can approach the user walking speed. These cases may reduce the accuracy of the classifier. Nevertheless, the different features help in reducing the ambiguity in this case as we quantify in Section~\ref{sec:evaluation}.

\subsection{Preprocessing}\label{Preprocessing}
The goal of this module is to reduce the noise in the input data, mainly due to the ping-pong effect. Specifically, due to the noisy nature of the wireless propagation and the unpredicted load on the cell tower, the user serving cell tower can change back and forth between different nearby serving cells. This is called the \emph{ping-pong} phenomenon \cite{henderson2008changing,song2004evaluating}. To handle this phenomenon, we apply a smoothing filter, where a low number of samples from a certain cell tower between two groups of samples from another dominant cell tower are replaced by the dominant cell ID. 

\subsection{Time and Frequency Domain Features}\label{sec:features}
 All features in \sys{} are calculated within a non-overlapping sliding window with a fixed size. Different parallel window sizes are used to capture different granularities. We combine the following features from both the time and frequency domain to enrich its features set:

\noindent \textbf{Time domain features:}

These features are extracted from the time domain for different window sizes covering changes in both the cell ID and RSS values.
\begin{enumerate}
\item Number of unique serving cell IDs within a time window: For a fixed window size, the higher this number, the higher the speed (Figure~\ref{fig:handoff}). This is directly related to the handoff frequency. 
\item Average cell residence time: Which is the average dwell time spent in each serving cell. The higher this number, the lower the phone speed (Figure~\ref{fig:ResisdenceTime}).

\item  RSS variance: Refers to the variance of the signal strength within the feature extraction window. This is based on the fact that higher speeds lead to faster changes and a noisier signal in the RSS and hence higher variance \cite{muthukrishnan2007sensing} (Figure~\ref{fig:rss_var}).

\item Average RSS difference between consecutive measurements: This is similar to the previous feature. In the extreme case, when the phone in stationary, one should expect that consecutive readings will almost be identical, which is complectly different from the high speed case (Figure~\ref{fig:AvgEUC}).

\end{enumerate}
\begin{table*}[!t]
\subfloat[Using RSS logarithmic scale]{
\scalebox{0.8}{
\renewcommand{\arraystretch}{2}
\begin{tabu}{ccc|[1pt]c|c|c|[1pt]}
\tabucline[1pt]{4-6}
&  & & \multicolumn{3}{ c|[1pt] }{\textbf{Predicted Movement}} \\ \cline{4-6}
&  & & Stationary & Walking & Driving \\ \tabucline[1pt]{2-6}

\multicolumn{1}{ c|[1pt] }{\multirow{3}{*}{\rotatebox[origin=c]{90}{\large \textbf{Precision}} }} &
\multicolumn{1}{ c }{\multirow{3}{*}{\rotatebox[origin=c]{90}{\textbf{Ground Truth}} }} &

\multicolumn{1}{ |c|[1pt] }{Stationary} & \cellcolor[gray]{0.9}\textbf{95.08} \% & 4.92 \% & 0.00 \% \\ \cline{3-6}
\multicolumn{1}{ c|[1pt]  }{}                 &        &

\multicolumn{1}{ |c|[1pt] }{Walking} & 0.00 \% & \cellcolor[gray]{0.9}\textbf{89.64} \% & 10.36 \%  \\ \cline{3-6}
\multicolumn{1}{ c|[1pt]  }{}                 &       &

\multicolumn{1}{ |c|[1pt] }{Driving} & 0.00 \% & 29.32 \% & \cellcolor[gray]{0.9}\textbf{70.67} \% \\ \tabucline[1pt]{2-6}
\end{tabu}}
\label{tab:PLog20}
}
\subfloat[Using RSS linear scale ]{
\scalebox{0.8}{
\renewcommand{\arraystretch}{2}
\begin{tabu}{ccc|[1pt]c|c|c|[1pt]}
\tabucline[1pt]{4-6}
&  & & \multicolumn{3}{ c|[1pt] }{\textbf{Predicted Movement}} \\ \cline{4-6}
&  & & Stationary & Walking & Driving \\ \tabucline[1pt]{2-6}

\multicolumn{1}{ c|[1pt] }{\multirow{3}{*}{\rotatebox[origin=c]{90}{\large \textbf{Precision}} }} &
\multicolumn{1}{ c }{\multirow{3}{*}{\rotatebox[origin=c]{90}{\textbf{Ground Truth}} }} &

\multicolumn{1}{ |c|[1pt] }{Stationary} & \cellcolor[gray]{0.9}\textbf{99.24} \% & 0.76 \% & 0.00 \% \\ \cline{3-6}
\multicolumn{1}{ c|[1pt]  }{}                 &        &

\multicolumn{1}{ |c|[1pt] }{Walking} & 0.00 \% & \cellcolor[gray]{0.9}\textbf{90.82} \% & 9.17 \%  \\ \cline{3-6}
\multicolumn{1}{ c|[1pt]  }{}                 &       &

\multicolumn{1}{ |c|[1pt] }{Driving} & 0.00 \% & 30.56 \% & \cellcolor[gray]{0.9}\textbf{69.43} \% \\ \tabucline[1pt]{2-6}
\end{tabu}}
}
\subfloat[Overall system (all features)]{
\scalebox{0.8}{
\renewcommand{\arraystretch}{2}
\begin{tabu}{ccc|[1pt]c|c|c|[1pt]}
\tabucline[1pt]{4-6}
&  & & \multicolumn{3}{ c|[1pt] }{\textbf{Predicted Movement}} \\ \cline{4-6}
&  & & Stationary & Walking & Driving \\ \tabucline[1pt]{2-6}

\multicolumn{1}{ c|[1pt] }{\multirow{3}{*}{\rotatebox[origin=c]{90}{\large \textbf{Precision}} }} &
\multicolumn{1}{ c }{\multirow{3}{*}{\rotatebox[origin=c]{90}{\textbf{Ground Truth}} }} &

\multicolumn{1}{ |c|[1pt] }{Stationary} & \cellcolor[gray]{0.9}\textbf{99.23} \% & 0.76 \% & 0.00 \% \\ \cline{3-6}
\multicolumn{1}{ c|[1pt]  }{}                 &        &

\multicolumn{1}{ |c|[1pt] }{Walking} & 0.00 \% & \cellcolor[gray]{0.9}\textbf{92.74} \% & 7.25 \%  \\ \cline{3-6}
\multicolumn{1}{ c|[1pt]  }{}                 &       &

\multicolumn{1}{ |c|[1pt] }{Driving} & 0.00 \% & 24.17 \% & \cellcolor[gray]{0.9}\textbf{75.82} \% \\ \tabucline[1pt]{2-6}
\end{tabu}}
}

\caption{Precision using features extracted from the RSS logarithmic scale, RSS linear scale, and both (overall system).}
\label{tab:log_res}
\end{table*}

\begin{table*}[!t]

\subfloat[Using RSS logarithmic scale]{
\scalebox{0.8}{
\renewcommand{\arraystretch}{2}
\begin{tabu}{ccc|[1pt]c|c|c|[1pt]}
\tabucline[1pt]{4-6}
&  & & \multicolumn{3}{ c|[1pt] }{\textbf{Predicted Movement}} \\ \cline{4-6}
&  & & Stationary & Walking & Driving \\ \tabucline[1pt]{2-6}

\multicolumn{1}{ c|[1pt] }{\multirow{3}{*}{\rotatebox[origin=c]{90}{\large \textbf{Recall}} }} &
\multicolumn{1}{ c }{\multirow{3}{*}{\rotatebox[origin=c]{90}{\textbf{Ground Truth}} }} &

\multicolumn{1}{ |c|[1pt] }{Stationary} & \cellcolor{LGreen}\textbf{100.00} \% & 0.00 \% & 0.00 \% \\ \cline{3-6}
\multicolumn{1}{ c|[1pt]  }{}                 &        &

\multicolumn{1}{ |c|[1pt] }{Walking} & 7.08 \% & \cellcolor{LGreen}\textbf{79.11} \% & 13.80 \%  \\ \cline{3-6}
\multicolumn{1}{ c|[1pt]  }{}                 &       &

\multicolumn{1}{ |c|[1pt] }{Driving} & 0.00 \% & 24.10 \% & \cellcolor{LGreen}\textbf{75.89} \% \\ \tabucline[1pt]{2-6}
\end{tabu}}
\label{tab:RLog20}
}
\subfloat[Using RSS linear scale]{
\scalebox{0.8}{
\renewcommand{\arraystretch}{2}
\begin{tabu}{ccc|[1pt]c|c|c|[1pt]}
\tabucline[1pt]{4-6}
&  & & \multicolumn{3}{ c|[1pt] }{\textbf{Predicted Movement}} \\ \cline{4-6}
&  & & Stationary & Walking & Driving \\ \tabucline[1pt]{2-6}

\multicolumn{1}{ c|[1pt] }{\multirow{3}{*}{\rotatebox[origin=c]{90}{\large \textbf{Recall}} }} &
\multicolumn{1}{ c }{\multirow{3}{*}{\rotatebox[origin=c]{90}{\textbf{Ground Truth}} }} &

\multicolumn{1}{ |c|[1pt] }{Stationary} & \cellcolor{LGreen}\textbf{100.00} \% & 0.00 \% & 0.00 \% \\ \cline{3-6}
\multicolumn{1}{ c|[1pt]  }{}                 &        &

\multicolumn{1}{ |c|[1pt] }{Walking} & 1.01 \% & \cellcolor{LGreen}\textbf{85.46} \% & 13.51 \%  \\ \cline{3-6}
\multicolumn{1}{ c|[1pt]  }{}                 &       &

\multicolumn{1}{ |c|[1pt] }{Driving} & 0.00 \% & 22.69 \% & \cellcolor{LGreen}\textbf{77.30} \% \\ \tabucline[1pt]{2-6}
\end{tabu}}
}
\subfloat[Overall system (all features)]{
\scalebox{0.8}{
\renewcommand{\arraystretch}{2}
\begin{tabu}{ccc|[1pt]c|c|c|[1pt]}
\tabucline[1pt]{4-6}
&  & & \multicolumn{3}{ c|[1pt] }{\textbf{Predicted Movement}} \\ \cline{4-6}
&  & & Stationary & Walking & Driving \\ \tabucline[1pt]{2-6}

\multicolumn{1}{ c|[1pt] }{\multirow{3}{*}{\rotatebox[origin=c]{90}{\large \textbf{Recall}} }} &
\multicolumn{1}{ c }{\multirow{3}{*}{\rotatebox[origin=c]{90}{\textbf{Ground Truth}} }} &

\multicolumn{1}{ |c|[1pt] }{Stationary} & \cellcolor{LGreen}\textbf{100.00} \% & 0.00 \% & 0.00 \% \\ \cline{3-6}
\multicolumn{1}{ c|[1pt]  }{}                 &        &

\multicolumn{1}{ |c|[1pt] }{Walking} & 1.00 \% & \cellcolor{LGreen}\textbf{87.76} \% & 11.22 \%  \\ \cline{3-6}
\multicolumn{1}{ c|[1pt]  }{}                 &       &

\multicolumn{1}{ |c|[1pt] }{Driving} & 0.00 \% & 18.24 \% & \cellcolor{LGreen}\textbf{81.76} \% \\ \tabucline[1pt]{2-6}
\end{tabu}}
}
\caption{Recall using features extracted from the RSS logarithmic, RSS linear scale, and both (overall system).}\label{tab:lin_res}
\end{table*}

\noindent \textbf{Frequency domain features:}
These features are extracted from the frequency domain using the fast Fourier transform algorithm (FFT) over different window sizes of RSS values.
\begin{enumerate}
\item Frequency with the highest energy: The intuition is that the movement speed affects the rate of RSS change. This can be captured by the dominant frequency after removing the DC component. 	
\item Signal energy: Defined as the sum of the square amplitudes of the FFT spectrum, which reflects the energy in the RSS signal. The slower the speed, the lower the energy in the signal should be. (Figure~\ref{fig:auto_corr}). 
    \begin{figure}[!t]
\centering
\includegraphics[width=0.6\linewidth]{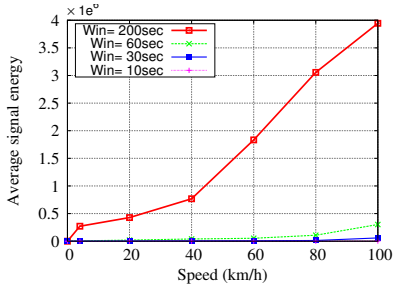}
\caption{Effect of phone speed on the signal energy.}
\label{fig:auto_corr}
\end{figure}

	\end{enumerate}
\subsection{Linear versus Logarithmic RSS Scale}
Typically, RSS is measured in the logarithmic scale. For example, using the free space loss model, the received signal strength in logarithmic scale ($p_r (\textrm{dB})$) at a distance $d$ from the transmitter is given by:
\begin{equation}
p_r (\textrm{dB})=p_0 -10*\alpha*\log\left( \frac{d}{d_0} \right)
\end{equation}
where $p_0$ is the power in dB at a reference distance $d_0$ from the transmitter and $\alpha$ is the path loss exponent.
Therefore, based on the RSS logarithmic scale, equal changes in the physical distance on the roads from the transmitter (i.e. serving cell tower) lead to equal distances in the RSS log space. However, this is not the case in the linear RSS scale, where equal road distances from the transmitter map to different distances in the linear RSS space. Based on this observation, \sys{} leverages both the logarithmic and linear RSS readings to extract the classification features. This scale diversity reduces the ambiguity between classes.

\subsection{Transportation Mode Classifier}\label{sec:classifier}

We use a decision tree classifier for differentiating between the three different modes of transportation due to its simplicity and efficient implementation . We use a total of 36 features (six main features from Section~\ref{sec:features} repeated for both linear and logarithmic scales for three different window sizes).

\subsection{Discussion}
\sys{} leverages features from different window sizes to enrich its feature space and hence obtain higher accuracy. As shown in this section, longer windows usually lead to better differentiation between the different speeds due to the more available information. However, longer windows increase the latency of estimation and extra large windows (not shown in this section) can worsen the performance as they may span different speeds within the same window.

Moreover, smaller window sizes can provide better differentiation in some features. For example, shorter window sizes can differentiate better between the stationary and walking modes for the cell resident time feature (Figure~\ref{fig:ResisdenceTime}).

To balance these factors, we use three window sizes in our implementation: 10, 30, and 60 seconds.

\section{Evaluation}\label{sec:evaluation}
In this section, we evaluate \sys{} in different real testbeds.We start by describing the experimental setup followed by the performance analysis.
\subsection{Experimental Setup}
We implemented a data collector on the Android Android SDK 2.3.3 (API Level 10). The collector gathers the serving cell tower ID, RSS, and time stamp periodically at a rate of one sample per second. The GPS location and speed are also collected as ground truth. The collector enables the user to add custom annotations, for example to manually enter the current ground truth transportation mode for traces that span multiple modes of transportation. We deployed our collector on different Android phones including a Samsung Nexus S, Nexus One, Galaxy S 7562, and  Galaxy Tab 1000. The collected traces were analyzed offline, were the features extraction and classifier were implemented using Matlab.

 using $135$ hours of cellular traces  collected by four users using different android

We collected a total of $135$ hours dataset over the course of eight months from January to August 2014 uniformly covering different modes of transportation. The combined traces length account for 485 km. Four different users were involved in the data collection process from the three operators in Egypt. 
\subsection{Performance Metrics}
We use five-fold cross validation to evaluate our classifier. We use two metrics for evaluation: precision and recall defined as:
\begin{itemize}
\item Precision= $\frac{\textrm{True positives}}{\textrm{True positives + False positives}}$= $\frac{\textrm{Correctly detected}}{\textrm{Total detected}}$.

    Precision captures the accuracy of the detected modes (percentage of detected modes that are correct).
\item Recall= $\frac{\textrm{True positives}}{\textrm{True positives + False negatives}}$= $\frac{\textrm{Correctly detected}}{\textrm{Total actual events}}$

\end{itemize}
High precision indicates that the classifier returns more correct results than incorrect ones. On the other hand, high recall means that the classifier detects most of the correct modes from the ground truth.

We use a confusion matrix between the different modes of transportation for each of the metrics. The values along the diagonal indicate the classifier performance. The off-diagonal elements quantify the confusion classes when error occurs. \subsection{Performance Analysis}\label{Panalysis}
In this section, we analyze the performance of the classifier for the different RSS scales using the different metrics. We start by showing the performance for the logarithmic and linear scales independently, followed by the combined classifier performance using all features (representing our overall system performance).  
\subsubsection{Features extracted from the log scale}
Table~\ref{tab:log_res},\ref{tab:lin_res} (a)  shows the confusion matrices for the precision and recall respectively. The table shows that the diversity of features from the time and frequency domain as well as features extracted from the different window sizes lead to high classification accuracy. The stationary mode is the easiest to detect. Most of the classification errors are between the walking and driving classes. This can be explained by noting that low driving speeds, e.g. due to traffic congestion, lead to speeds that are comparable to the walking speeds. The driving mode of transportation is the hardest due to the wide span of speeds it covers compared to the other modes. Nevertheless, the average precision and recall are 85.13\% and 85\% respectively. These are further enhanced by combining the logarithmic and linear features as we quantify in Section~\ref{sec:combined_res}.

\subsubsection{Features extracted from the linear scale}
Table~\ref{tab:log_res}, \ref{tab:lin_res} (b) shows the confusion matrices for the precision and recall respectively based on the features extracted from the RSS linear scale. The table shows similar results to logarithmic scale results. Linear scale features slightly perform better for the stationary and walking classes while the logarithmic scale features perform better in the driving class. This validates our observation that the features are independent and can be fused together to provide better performance as in the next section.

\subsubsection{Overall system performance: Combined log and linear space features}
\label{sec:combined_res}
Table~\ref{tab:log_res},\ref{tab:lin_res}(c)  shows the confusion matrices using all the 36 combined features. The table shows that using all features lead to an overall average precision and recall of 89.26\% and 89.84\% respectively. This highlights that \sys{} can achieve its goals of ubiquitous, high accuracy, and energy-efficient transportation mode detection using minimal information from the serving cell tower only.

\section{Conclusion}\label{Sec:conclusion}
We presented a ubiquitous, energy-efficient, and accurate mode of transportation detection system, \sys{}, that completely relies on the information from the serving cell tower only. We showed how \sys{} leverages the diversity of features in the logarithmic and linear scales, time and frequency domain, as well as different window sizes to extend its feature space and achieve high accuracy.

Real world implementation and experiments using different Android phones spanning 135 hours and 485 km over an eight months period confirm the effectiveness of \sys{} showing an average precision and recall of 89.26\% and 89.84\% respectively, highlighting \sys{} ability meet its goals of high accuracy, energy efficiency, and ubiquitous deployment on different phone types as well as on the user and cellular provider sides.

Currently, we are expanding \sys{} in different directions including leveraging more features, classifying more modes of transportation, implementation on other phone operating systems, among others.

\section*{Acknowledgment}
This work is supported in part by the Information Technology Industry Development Agency (ITIDA).

\bibliographystyle{IEEEtran}
\bibliography{Collection.bbl}

\end{document}